\documentclass[runningheads]{llncs}

\include{psd-macros}
\usepackage[linesnumbered,boxed,lined,commentsnumbered,algochapter]{algorithm2e}
\usepackage{subfigure}
\usepackage{times}
\usepackage[toc,page]{appendix}

\title{NeuralExplorer: State Space Exploration of Closed Loop Control Systems Using Neural Networks}

\author{%
  Manish Goyal 
  \and
 Parasara Sridhar Duggirala \\ 
 \texttt{\{manishg,psd\}@cs.unc.edu}
} 
\authorrunning{M. Goyal, P. S. Duggirala}

\institute{Department of Computer Science, University of North Carolina at Chapel Hill
}

\begin{document}

\maketitle

\begin{abstract}%
In this paper, we propose a framework for performing state space exploration of closed loop control systems. 
Our approach involves approximating sensitivity and a newly introduced notion of inverse sensitivity by a neural network.
We show how the approximation of sensitivity and inverse sensitivity can be used for computing estimates of the reachable set.
We then outline algorithms for performing state space exploration by generating trajectories that reach a neighborhood.
We demonstrate the effectiveness of our approach by applying it not only to standard linear and nonlinear dynamical systems, but also to nonlinear hybrid systems and also neural network based feedback control systems. 
\end{abstract}


\section{Introduction}
\label{sec:intro}
The decrease in the cost of embedded microcontrollers and integration of software into many control processes has lead to the deployment of increasingly sophisticated control algorithms in many safety critical scenarios. 
Additionally, the recent improvements in software and hardware platforms for training and evaluation of neural networks have made it easier to integrate them in embedded devices.
This increased complexity of software in control tasks makes testing and validation of the closed loop systems very challenging. 

Testing is the most commonly used technique for checking whether a system satisfies its specification. 
In a typical work flow, after designing the feedback function, the control designer generates a few test cases for the closed loop system and checks if they satisfy the specification. 
Given that the state space is continuous, finding the trajectory that violates the specification is similar to searching for a needle in a haystack. 
For example, consider a regulation application where the output of the control system is stabilized to a set point $s$ with the error threshold of $\delta$. Therefore, the output should remain in the interval $[s-\delta, s+\delta]$. If all the test cases satisfy the specification, the control designer does not know the next test case input that results in a higher value of error than observed in the test cases.

While the designer can encode the property as a temporal logic formula and use off-the-shelf falsification tools, such an approach has some drawbacks. 
First, falsification tools are geared towards finding a trajectory that violates the given specification, not necessarily to help the control designer in state space exploration. 
Second, if the threshold of $\delta$ is increased, the results from the falsification analysis are no longer useful. 
Finally, using falsification tools would require the specification to be given in a temporal logic such as signal temporal logic or metric temporal logic. 
While such specification might be useful in the verification and deployment phase, they are sometimes a hinderance during the design phase. 
What the control designer needs, is a tool that helps her to systematically generate trajectories that result in various values of error thresholds. 
Currently, there are no tools for performing state space exploration based on the previous tests generated or based on the input given by the control designer.

In this paper, we present \emph{NeuralExplorer}, a technique for performing state space exploration of closed loop control systems using neural networks. 
After the control designer generates a few test scenarios, she can use NeuralExplorer to generate test case scenarios that reach a neighborhood. 
%
%Currently, there are not tools at the disposal of control designer to automatically search for trajectories that move closer to the boundary values of $s-\delta$ and $s+\delta$. %
%While NeuralExplorer might not always yield a trajectory that reaches the target state, it helps the control designer to perform systematic exploration of the state space.
%
%Our goal is to equip the control engineer with a mechanism for performing state space exploration. Often, control engineers generate a few test simulations. If these test simulations do not turn out to be unsafe, then more test simulations are generated. Currently, there is no way to use the information from the generated from the sample tests generated to help the control engineer to generate new test cases.
%
%In this paper, we present NeuralExplorer, a framework for performing state space exploration of close loop control systems using Neural Networks. 
%
The artifact that helps us in this endeavor is \emph{sensitivity}. 
Informally, sensitivity of a closed loop system is the change in the trajectory of the system as a result of perturbing the initial condition. 
In this paper, we also present a backward time notion of sensitivity called \emph{inverse sensitivity}. 
Given a sample set of trajectories in the domain of interest, we train a neural network to learn the sensitivity and the inverse sensitivity functions. 
This neural network is then used to generate a trajectory (or trajectories) that reaches a destination (or a neighborhood around it). 

Our framework has two primary advantages. 
First, since NeuralExplorer relies only on the sample test cases, it does not require a model of the system and can be applied to a black-box systems. 
Second, since sensitivity is a fundamental property of the closed loop system, approximating it using a neural network is generalizable to trajectories that are beyond the test cases generated by the control designer. 
%Our goal in developing NeuralExplorer is to equip the control designer to perform a systematic exploration of state space and employ tools like reachability when the thresh
In evaluating our framework, we were not only able to perform state space exploration for standard linear and nonlinear dynamical systems, but also for nonlinear hybrid systems and neural network based feedback control systems.
We believe that NeuralExplorer is useful for generating corner cases and supplements some of the existing testing and reachable set computation procedures.

%Such a technique would be helpful for a control designer to perform directed state space exploration of closed loop control systems.

% \begin{itemize}
%   \item Limit the main text (not counting references) to 8 PMLR-formatted pages, using this template.
%   \item Include {\em in the main text} enough details, including proof details, to convince the reviewers of the contribution, novelty and significance of the submissions.
% \end{itemize}

\section{Related Work}

\textit{Reachability} analysis is often employed for proving the safety specification of safety critical control system~\cite{10.1007/3-540-64358-3_34,10.1007/3-540-36580-X_4}. Some of the recent works in this domain are SpaceEx~\cite{spaceex}, Flow*~\cite{10.1007/978-3-642-39799-8_18}, CORA~\cite{althoff2015introduction} and HyLAA~\cite{bak2017hylaa}. These techniques use a symbolic representation for the reachable set of states. While these are useful for proving that the safety specification is satisfied, generating counterexamples using reachability analysis is still an area of research~\cite{DBLP:conf/adhs/GoyalD18,GOYAL2020109005}. 

For generating trajectories that violate a given safety specification, falsification techniques are widely applied~\cite{Fainekos09,DonzeM10}. In these techniques, the required specification is specified in a temporal logic such as Metric Temporal Logic (MTL)~\cite{MTLmain} or Signal Temporal Logic (STL)~\cite{STLmain,Kyriakis:2019:SMR:3365919.3358231}. Given a specification, falsification techniques generate several sample trajectories and use various heuristics~\cite{MCFalsification,AbbasF11,SankaranarayananF12,zutshi2014multiple,DeshmukhFKS0J15,ghosh2016diagnosis} for generating trajectories that violate the specification. Prominent tools in this domain include S-Taliro~\cite{AnnpureddyLFS11} and Breach~\cite{BreachAD}.

Bridging falsification and reachability are simulation driven verification methods~\cite{donze2007systematic,duggirala2013verification,huang2014proofs,fan2015bounded}. These methods compute an upper bound on the sensitivity of the trajectories and compute an over-approximation of the reachable set using sample trajectories. While these techniques bridge the gap between falsification and verification, they suffer from curse of dimensionality. That is, the number of trajectories generated might increase exponentially with the dimensions. C2E2~\cite{DBLP:conf/hybrid/DuggiralaPM015}, and DryVR~\cite{DBLP:conf/cav/FanQM017} are some of the well known tools in this domain. 

Given the rich history of application of neural networks in control~\cite{miller1995neural,lewis1998neural,moore2012iterative} and the recent advances in software and hardware platforms, neural networks are now being deployed in various control tasks. As a result, many verification techniques are now being developed for neural network based control systems~\cite{ivanov2018verisig,Tran:2019:SVC:3365919.3358230,Sun:2019:FVN:3302504.3311802,DBLP:conf/hybrid/DuttaCJST19}. Additionally, techniques for verification of neural networks deployed in other domains have been proposed~\cite{DBLP:conf/iclr/TjengXT19,DBLP:journals/tecs/SunHKSHA19,DBLP:conf/cav/HuangKWW17}. 

In this paper, we use neural networks to approximate an underlying property of sensitivity. We believe that this is a valid approach because recently, many neural network based frameworks for learning the dynamics or their properties have been proposed~\cite{DBLP:journals/complexity/PanD18,PhysRevLett.120.024102,pmlr-v80-long18a,raissi2018multistep,chen2018neural,Phan:2019:NSC:3313149.3313372}.

% for ,  use symbolic representations for the reachable set, while  use system simulations. \textit{Falsification} ~\cite{Fainekos09,DonzeM10} is orthogonal to verification. In addition to the tools S-Taliro~\cite{AnnpureddyLFS11} and Breach~\cite{BreachAD}, other falsification techniques employ a variety of methods~\cite{MCFalsification,AbbasF11,SankaranarayananF12,zutshi2014multiple,DeshmukhFKS0J15,ghosh2016diagnosis} for discovering an execution that violates the specification given in  ~\cite{DBLP:conf/adhs/GoyalD18} provides a technique for classifying violating executions based on various metrics.

% The presented work is aligned with neural network based learning and verification/falsification. Neural networks are extensively used to predict controllers~\cite{miller1995neural, ivanov2018verisig, Tran:2019:SVC:3365919.3358230}, perform state classification , and learn system dynamics from data~\cite{DBLP:journals/complexity/PanD18, PhysRevLett.120.024102, pmlr-v80-long18a, raissi2018multistep, chen2018neural}. Given their vast applicability, another body of work is focused on exploring verification techniques for neural networks.
\section{Preliminaries}
\label{sec:prelims}

We denote the elements of the state space as $x$ to be elements in $\reals^{n}$. Vectors are denoted as $v$. We denote the dynamics of the plant as 

\begin{equation}
\label{eq:sys}
    \dot{x} = f(x,u)
\end{equation}

Where $x$ is the state space of the system that evolves in $\reals^n$ and $u$ is the input space in $\reals^m$. 

% \begin{definition}[Trajectories of Open Loop System]
% \label{def:trajectory}
% Given an input signal $u: \nnreals \rightarrow \reals^{m}$, initial state $x_0$, and time $t \geq 0$, the trajectory of the system $\xi(x_0, u, t)$ provides the state of the system after time $t$.
% \end{definition}

\begin{definition}[Unique Trajectory Feedback Functions]
\label{def:utfeedback}
A feedback function $u = g(x)$ is said to be unique trajectory feedback function if the closed loop system $\dot{x} = f(x,g(x))$ is guaranteed existence and uniqueness of the solution for the initial value problem for all initial points $x_0 \in \reals^n$.
\end{definition}

Notice that for a feedback function to give a unique trajectory feedback, it need not be differentiable. From the sufficuent conditions of ODE solutions, it is sufficient if $g(x)$ is continuous and is lipschitz. 

\begin{definition}[Trajectories of Closed Loop System]
\label{def:cltraj}
Given a unique trajectory feedback function $u=g(x)$, a trajectory of closed loop system $\dot{x} = f(x,g(x))$, denoted as $\xi_{g}(x_0,t)$ ($t \geq 0$), is the solution of the initial value problem of the differential equation $\dot{x} = f(x,g(x))$ with initial condition $x_0$. We often drop the feedback function $g$ when it is clear from the context.

We extend the notion of trajectory to include backward time trajectories as well. Given $t > 0$, the backward time trajectory $\xi_{g}(x_0, -t) = x$ such that $\xi_{g}(x,t) = x_0$. We denote backward time trajectory as $\xi^{-1}(x_0,t)$.
\end{definition}

Given $x_0, x_1 \in \reals^n$  and $t > 0$ such that $\xi(x_0,t) = x_1$, then $\xi^{-1}(x_1,t) = x_0$. It is trivial to observe that $\xi(\xi^{-1}(x_0,t),t) = x_0$.

\begin{definition}[Sensitivity of Trajectories]
\label{def:sensitivity}
Given an initial state $x_0$, vector $v$, and time $t$, the sensitivity of the trajectories, denoted as $\Phi(x_0, v, t)$ is defined as.
$$
\Phi(x_0,v,t) = \xi(x_0+v,t) - \xi(x_0,t).
$$
Informally, sensitivity is the vector difference between the trajectories starting from $x_0$ and $x_0 + v$ after time $t$. We extend the definition of sensitivity to backward time trajectories as 
$$
\Phi^{-1}(x_0,v,t) = \xi^{-1}(x_0+v,t) - \xi^{-1}(x_0,t).
$$
\end{definition}

We call $\Phi^{-1}(x_0,v,t)$ as inverse sensitivity function. 
Informally, inverse sensitivity function gives us the perturbation of the initial condition that is required to displace the trajectory passing through $x_0$ by $v$.
Observe that $\xi(\xi^{-1}(x_0,t)+\Phi^{-1}(x_0,v,t), t) = x_0 + v$. 
%Informally, $\Phi^{-1}$ tells us the change in the initial state of trajectory required to achieve the displacement of $v$ at a point of time in future of the trajectory.

For general nonlinear differential equations, analytic representation of the trajectories of the ODEs need not exist. If the closed loop system is a smooth function, then the infinite series for the trajectories is given as
\begin{equation}
\xi(x_0,t) = x_0 + \L_{f}(x_0)t + \L_{f}^2(x_0)\frac{t^{2}}{2!} + \L_{f}^{3}(x_0)\frac{t^3}{3!} + \ldots.
\label{eq:infSol}    
\end{equation}

Where $\L_{f}^{i}$ is the $i^{th}$ order Lie-derivative over the field $f(x,g(x))$ at the state $x_0$. Hence, one can write the sensitivity function as 
\begin{equation}
\Phi(x_0,v,t) = v + (\L_{f}(x_0+v)-\L_{f}(x_0))t + (\L_{f}^2(x_0+v)-\L_{f}^2(x_0))\frac{t^2}{2!} + \ldots.
\label{eq:infSen}    
\end{equation}

$\Phi^{-1}(x_0,v,t)$ is obtained by substituting $-f$ for $f$ in Equation~\ref{eq:infSen}. When the closed loop dynamics is linear, i.e., $\dot{x} = Ax$, it is easy to observe that $\Phi(x_0,v,t) = e^{At}v$, $\Phi^{-1}(x_0,v,t) = e^{-At}v$ where $e^{At}$ ($e^{-At}$) is the matrix exponential of the matrix $At$ ($-At$). Obseve that for linear systems, the inverse sensitivity function is independent of the state $x_0$.  

For nonlinear dynamical systems, one can truncate the infinite series upto a specific order and obtain an approximation. However, for hybrid systems that have state based mode switches, or for feedback functions where the closed loop dynamics is not smooth or is discontinuous, such an infinite series expansion is hard to compute. The central idea in this paper is to approximate  $\Phi$ and $\Phi^{-1}$ using a neural network and perform state space exploration using such neural networks.
\section{Neural Network Approximations of Sensitivity and Inverse Sensitivity}
\label{sec:approxinvsen}

%In this section, we outline the approach for approximating sensitivity and inverse sensitivity by a neural network. 

%\subsection{Approximating Inverse Sensitivity Using Neural Networks}

Given a domain of operation $D \subseteq \reals^n$, one can generate a finite set of trajectories for testing the system operation in $D$. 
Often, these trajectories are generated using numerical ODE solvers which return trajectories sampled at a regular time step. 
For approximating sensitivity and inverse sensitivity, we generate a finite number of time bounded trajectories where the step size, time bound, and the number of trajectories are specified by the user. 
The trajectories can be generated either according to a probability distribution specified by the user or from specific initial configurations provided by her.

Given a sampling of a trajectory at regular time interval with step size $h$, i.e., $\xi(x_0,0)$, $\xi(x_0,h)$, $\xi(x_0,2h)$, $\ldots$, $\xi(x_0, kh)$, we make two observations. 
First, any prefix or suffix of this sequence is also a trajectory, albeit, of a shorter duration. Hence, from a given set of trajectories, one can generate more \emph{virtual trajectories} that have shorter duration. 
Second, given two trajectories (real or virtual) starting from initial states $x_1$ and $x_2$, ($x_1 \neq x_2$), We have the two following observations.
\begin{eqnarray}
\Phi(x_1, x_2-x_1, t) &=& \xi(x_2,t) - \xi(x_1,t) \label{eq:supplement1} \\
\Phi^{-1}(\xi(x_1,t), \xi(x_2,t) - \xi(x_1,t), t) &=& x_2 - x_1. \label{eq:supplement2}
\end{eqnarray}

Given an initial set of trajectories, we generate virtual trajectories and use Equations~\ref{eq:supplement1} and~\ref{eq:supplement2} for generating all tuples $\langle x_0, v, t, v_{sen} \rangle$ and $\langle x_0, v, t, v_{isen}\rangle$ such that $v_{sen} = \Phi(x_0, v, t)$ and $v_{isen} = \Phi^{-1}(x_0, v, t)$. 
This data is then used for training and evaluation of the neural network to approximate the inverse sensitivity function. 
We denote these networks as $\mathit{NN}_{\Phi}$ and $\mathit{NN}_{\Phi^{-1}}$ respectively.
%From these two observations and given the initial set of trajectories, we generate all pairs of 

\subsection{Evaluation on Standard Benchmarks}
\label{sec:trainingNNs}

For approximating the sensitivity and inverse sensitivity functions, we pick a standard set of benchmakrs consisting of nonlinear dynamical systems, hybrid systems, and control systems with neural network feedback functions. 
Most of the benchmarks are taken from standard hybrid systems benchmark suite~\cite{benchmarksurl,DBLP:conf/adhs/ImmlerA0FFKLMTZ18,Bak2019}.
The benchmarks \texttt{Brussellator}, \texttt{Lotka}, 
\texttt{Jetengine}, \texttt{Buckling}, 
\texttt{Vanderpol}, \texttt{Lacoperon}, \texttt{Roesseler}, \texttt{Steam}, \texttt{Lorentz}, %\texttt{LaubLoomis} 
and \texttt{Coupled vanderpol} are continuous nonlinear systems, where \texttt{Lorentz} and \texttt{Roesseler} are \textit{chaotic} as well.  \texttt{SmoothHybrid-} \texttt{-Oscillator} and \texttt{HybridOscillator} are nonlinear hybrid systems.
%; \texttt{Vehicle Platoon} is a linear hybrid system with 2 discrete locations. 
%
The remaining benchmarks \texttt{Mountain Car} and \texttt{Quadrotor} are selected from~\cite{ivanov2018verisig}, where the state feedback controller is given in the from of neural network. 
%
% The hybrid dynamical systems, \textit{Quadrotor} 
% and \texttt{Vehicle Platoon} are 
% is made deterministic such that their transitions have urgent semantics~\cite{Minopoli:2016:SMH:2883817.2883825} that is, a transition is taken as soon as the corresponding guard is enable. 
%
% Particularly in \textit{Quadrotor} benchmark, we induce determinism by fixing the control law based on 8 control actions generated by the controller.

% \subsection{Training and Testing}

For each benchmark, we generated a given number ($N$ is typically 30 or 50) of trajectories, where the step size for ODE solvers ($s$) and the time bound are provided by the user. 
The data used for training the neural network is collected as described in~\ref{sec:approxinvsen}. We use 90\% of the data for training and 10\% for testing.
%
%for a desired step size ($s$) and time bound ($|\xi|$) which are provided in Table~\ref{tab:results}. 
%For all pairs of points $x_0$ and $x_1 = (x_0 + v)$ at time instances $t = kT$, where $ \forall T \in \{1,2,4,8,10,12,16\}, 1 \leq k \leq |\xi|\bmod T$, in any two trajectories $\xi_1$ and $\xi_2$, we train the perturbation function as  $\hat{\Phi}^{-1}(x_0, x_1-x_0, t) = \xi^{-1}_{2}(x_1,t) - \xi^{-1}_{1}(x_0,t)$. 
%90\% of samples for are used for training and 10\% for testing. 
We used the Python Multilayer Perceptron implemented as Sequential model in \texttt{Keras}~\cite{chollet2015keras} library with Tensorflow as the backend. The network has 8 layers with each layer having 512 neurons. 
The optimizer used is stochastic gradient descent. 
The network is trained using Levenberg-Marquardt backpropagation algorithm optimizing the mean absolute error loss function, and the Nguyen-Widrow initialization. 

The activation function used to train the network is \textbf{relu} for all benchmarks except \texttt{Mountain car} for which \textbf{sigmoid} performs better because the NN controller is sigmoid-based. 
Note that the choice of hyper-parameters such as number of layers and neurons, the loss and activation functions is empirical, and is motivated by some prior work~\cite{darsgoyalpsd2019}. 
We evaluate the network performance using root mean square error (MSE)
%$\Big(\frac{1}{n}\sum_{i=1}^{n}{({\Phi}^{-1}(x,v,t)-\hat{\Phi}^{-1}(x,v,t))^2}\Big)$ 
and mean relative error (MRE) 
%$\Big(\frac{1}{n}\sum_{i=1}^{n}{\Big(\frac{\left\lVert{\Phi}^{-1}(x,v,t)-\hat{\Phi}^{-1}(x,v,t)\right\rVert_2}{\left\lVert{\Phi}^{-1}(x,v,t)\right\rVert_2}\Big)}\Big)$ 
metrics.
The training and evaluation are performed on a system running Ubuntu 18.04 with a 2.20GHz Intel Core i7-8750H CPU with 12 cores and 32 GB RAM. 
The network training time, MSE and MRE are given in
%columns 6, 7 and 8, respectively, in
Tables~\ref{tab:training-forward-sensitivity} and~\ref{tab:training-inverse-sensitivity} respectively.

\begin{table}[ht!]
\footnotesize
\centering
\begin{tabular}{|*{8}{c|} }
\hline
\multicolumn{2}{|c|}{Benchmark} & Dims & Step size & Time bound & NN Training & MSE & MRE \\
%\cline{1-8}
& & & (sec) & & Time (min) & $~$ & $~$ \\
\hline
\hline
\multirow{11}{*} & Brussellator & 2 & 0.01 & 500 & 40.0 & 0.14 & 0.34\\
\cline{2-8}
& Buckling & 2 & 0.01 & 500 & 25.0 & 2.38 & 0.18\\
\cline{2-8}
& Lotka & 2 & 0.01 & 500 & 27.0 & 0.38 & 0.31\\
\cline{2-8}
& Jetengine & 2 & 0.02 & 300 & 35.0 & 0.086 & 0.63\\
\cline{2-8}
Continuous & Vanderpol & 2 & 0.01 & 500 & 75.50 & 0.15 & 0.29\\
\cline{2-8}
Nonlinear & Lacoperon & 2 & 0.1 & 500 & 66.0 & 0.12 & 0.33\\
\cline{2-8}
Dynamics & Roesseler & 3 & 0.02 & 500 & 42.0 & 0.58 & 0.087\\
\cline{2-8}
 & Lorentz & 3 & 0.01 & 500 & 22.0 & 1.08 & 0.11\\
\cline{2-8}
& Steam & 3 & 0.01 & 500 & 35.0 & 0.34 & 0.07\\
\cline{2-8}
& C-Vanderpol & 4 & 0.01 & 500 & 70.0 & 0.18 & 0.15\\
\hline
\multirow{5}{*} & HybridOsc. & 2 & 0.01 & 500 & 80.0 & 0.35 & 0.11\\
\cline{2-8}
Hybrid/ & SmoothOsc. & 2 & 0.01 & 500 & 38.5 & 0.40 & 0.096\\
\cline{2-8}
NN Systems & Mountain Car & 2 & - & 100 & 12.5 & 0.015 & 0.79\\
\cline{2-8}
 & Quadrotor & 6 & 0.01 & 120 & 40.0 & 0.064 & 0.20\\
\hline
\end{tabular}
\caption{\textbf{Learning senstivity function.} The set of benchmarks includes nonlinear dynamical, hybrid and neural network based feedback control systems. Time bound is number of steps for which the system  simulation is computed.
}
\label{tab:training-forward-sensitivity}
% \end{table}

% \begin{table}[ht!]
\footnotesize
\centering
\begin{tabular}{|*{8}{c|} }
\hline
\multicolumn{2}{|c|}{Benchmark} & Dims & Step size & Time bound & NN Training & MSE & MRE \\
%\cline{1-8}
& & & (sec) & & Time (min) & $~$ & $~$ \\
\hline
\hline
\multirow{11}{*} & Brussellator & 2 & 0.01 & 500 & 67.0 & 1.01 & 0.29\\
\cline{2-8}
& Buckling & 2 & 0.01 & 500 & 42.0 & 0.59 & 0.17\\
\cline{2-8}
& Lotka & 2 & 0.01 & 500 & 40.0 & 0.50 & 0.13\\
\cline{2-8}
& Jetengine & 2 & 0.01 & 300 & 34.0 & 1.002 & 0.26\\
\cline{2-8}
& Vanderpol & 2 & 0.01 & 500 & 45.50 & 0.23 & 0.23\\
\cline{2-8}
Continuous & Lacoperon & 2 & 0.2 & 500 & 110.0 & 1.8 & 0.46\\
\cline{2-8}
Nonlinear & Roesseler & 3 & 0.02 & 500 & 115.0 & 0.44 & 0.07\\
\cline{2-8}
Dynamics & Lorentz & 3 & 0.01 & 500 & 67.0 & 0.48 & 0.08\\
\cline{2-8}
& Steam & 3 & 0.01 & 500 & 58.0 & 0.13 & 0.057\\
%\cline{2-8}
%& Rober & 3 & 0.1 & 500 & 42.0 & 0.24 & 0.044\\
\cline{2-8}
& C-Vanderpol & 4 & 0.01 & 500 & 75.0 & 0.34 & 0.16\\
%\cline{2-8}
%& LaubLoomis & 7 & 0.1 & 200 & 103.0 & 0.15 & 0.50\\
\hline
\multirow{5}{*} & HybridOsc. & 2 & 0.01 & 1000 & 77.0 & 0.31 & 0.077\\
\cline{2-8}
Hybrid/ & SmoothOsc. & 2 & 0.01 & 1000 & 77.5 & 0.23 & 0.063\\
\cline{2-8}
NN Systems & Mountain Car & 2 & - & 100 & 10.0 & 0.005 & 0.70\\
\cline{2-8}
 & Quadrotor & 6 & 0.01 & 120 & 25.0 & 0.0011 & 0.16\\
% \cline{2-8}
%  & VehiclePlatoon & 9 & 0.02 & 500 & 121.0 & 0.071 & 0.20\\
\hline
\end{tabular}
\caption{\textbf{Learning inverse senstivity function.} Parameters and performance of neural network tasked with learning inverse sensitivity function.
% Both of the Hybrid Oscillator benchmarks originally have 2 modes. The numeral suffix (1 or 2) in their names represents that the corresponding hybrid system has either first mode or second. 
}
\label{tab:training-inverse-sensitivity}
\end{table}

\section{Space Space Exploration Using Neural Network Approximation}
\label{sec:spaceexplore}
%\subsection{Space Space Exploration Using Neural Network Approximation}
\label{sec:neuralexplore}

In this section, we present various applications in the domain of state space exploration using the neural network approximations of sensitivity and inverse sensitivity. 
The goal of state space exploration is to search for trajectories that satisfy or violate a given specification. 
In this paper, we primarily concern ourselves with a safety specification, that is, whether a specific trajectory reaches a set of states labelled as \emph{unsafe}.
In order to search for such trajectories, we present various empirical techniques that use both forward and inverse sensitivity.

\subsection{Reaching a Specified Destination Using Inverse Sensitivity Approximation}
\label{sec:reachgoal}

In the course of state space exploration, after testing the behavior of the system for a given set of test cases, the control designer might choose to explore the behavior of a system that reaches a destination or approaches the boundary condition for safe operation. Given a domain of operation $D$, we assume that the designer provides a desired target state $z$ (with an error threshold of $\epsilon$) that is reached by a trajectory at time $t$. Our goal is to generate a trajectory $\xi$ such that $\xi(t)$ visits a state in the $\epsilon$ neighborhood of the target $z$.

Our approach for generating the target trajectory is as follows. First, we generate a random trajectory $\xi$ from the initial set $\Theta$, and compute the difference vector of target state $z$ and $\xi(t)$. We now use the neural network approximation of the inverse sensitivity function and estimate the perturbtion required in the initial set such that the trajectory after time $t$ goes through $z$. Since the neural network can only approximate the inverse sensitivity function, the trajectory after the perturbation need not visit $\epsilon$ neighborhood of the destination. However, we can repeat the procedure until a threhold on the number of iterations is reached or the $\epsilon$ threshold is satisfied. The pseudocode of this procedure, denoted as \textsf{reachTarget}, is given in Algorithm~\ref{alg:basicapproach}

\begin{algorithm}[H]
\SetAlgoLined
\SetKwInOut{Input}{input}\SetKwInOut{Output}{output}\SetKw{Return}{return}
\Input{System simulator $\xi$, Time bound $T$, trained neural network $\mathit{NN}_{\Phi^{-1}}$, time instance $t \leq T$, target state $z \in D$, iteration count $I$, initial set $\theta$, and threshold $\epsilon$}
\Output{State $x \in \theta$, $d_r$, $d_a \deq ||\xi(x_0, t) - z||_2$ such that $d_a \leq \epsilon$}
 $x \gets x_{random} \in \theta$; $i \gets 1$\; 
 $x_0 \gets \xi(x, t)$; $d_a \gets ||x - z||_2$\; $d_{orig} \gets d_a$; $d_r \gets 1$\; \label{ln:initialize}
 \While{$(d_a > \epsilon$) \& ($i \leq I)$ }{\label{ln:begLoop1}
  $v \gets x_0 - z$\; 
  $v^{-1} \gets \mathit{NN}_{\Phi^{-1}}(x_0, v, t)$\;
  $x \gets x + v^{-1}$; $x_0 \gets \xi(x, t)$\;
  $d_a \gets ||x_0 - z||_2$; $i \gets i + 1$\;
 }\label{ln:endLoop1}
 $d_r \gets \frac{d_a}{d_{orig}}$\;
 {\bf return} $(x, d_r, d_a)$\;
 \caption{\textsf{reachTarget}. Finding an initial state from which the simulation goes within $\epsilon$-neighborhood of destination $z$ at time $t$.}
 \label{alg:basicapproach}
\end{algorithm}

%We now explain the state space exploration algorithm, \textsf{reachTarget}, as given in Algorithm~\ref{alg:basicapproach}. 
%Given a target state $z$ and time instance $t$, it tries to find an initial state $x_0$ such that the simulation starting from $x_0$ goes as close as possible to $z$ at time $t$. 
We initialize the variables in line~\ref{ln:initialize}. $\xi(x,.)$ is the simulation generated by $\xi$ for the state $x$. The \textbf{while} loop (lines~\ref{ln:begLoop1}-~\ref{ln:endLoop1}) runs until either the threshold is reached or iteration count is exhausted. $v^{-1} \deq \Phi^{-1}(x_0,v,t)$ is the inverse sensitivity function which is learned using a neural network $\mathit{NN}_{{\Phi}^-1}$; $\left\lVert\cdot\right\rVert_2$ is the $l2$-norm. 
%The algorithm is illustrated using \texttt{Steam} and \texttt{Rober} benchmarks in Figure~\ref{fig:base_approach_exp}.

\subsubsection{Evaluation of \textsf{reachTarget} on Standard Benchmarks}

\begin{table}[H]
\footnotesize
\centering
\begin{tabular}{|*{8}{c|} }
\hline
Benchmark & Dims & \multicolumn{3}{|c|}{Iteration count = 1} & \multicolumn{3}{|c|}{Iteration count = 5} \\
\cline{3-8}
& & $d_a$ & $d_{r}$ & Time (ms) & $d_a$ & $d_{r}$& Time (ms)  \\
\hline
\hline
Brussellator & 2 &  [0.19 - 1.87] & [0.23 - 0.74] & 11.38 & [0.003- 0.22] & [0.01 - 0.12] & 31.34\\
\hline
Buckling & 2 &  [1.67 - 11.52 & [0.17 - 0.45] & 13.61 & [0.36- 2.09] & [0.06 - 0.31] & 34.51\\
\hline
Lotka & 2 & [0.08 - 0.24] & [0.21 - 0.45] & 12.38 & [0.02 - 0.07] & [0.09 - 0.22] & 34.28\\
\hline
Jetengine & 2 & [0.05 -0.20] & [0.19 - 0.28] & 15.96 & [0.0004 - 0.05] & [0.006 - 0.14] & 38.26\\
\hline
Vanderpol & 2 & [0.29 - 0.58] & [0.16 - 0.66] & 12.34 & [0.03 - 0.18] & [0.04 - 0.16] & 34.02\\
\hline
Lacoperon & 2 & [0.03 - 0.13] & [0.12 - 0.28] & 17.18 & [0.003 - 0.03] & [0.02 - 0.16] & 37.34\\
\hline
Roesseler & 3 & [0.72 - 2.02] & [0.20 - 0.34] & 16.08 & [0.21 - 0.63] & [0.06 - 0.14] & 38.26\\
\hline
Lorentz & 3 & [1.24 - 5.60] & [0.29 - 0.58] & 24.72 & [0.20 - 0.70] & [0.05 - 0.17] & 60.18\\
\hline
Steam & 3 & [1.59 - 5.21] & [0.31 - 0.67] & 8.68 & [0.41 - 1.8] & [0.08 - 0.30] & 69.80\\
% \hline
% Rober & 3 & [1.01 - 2.37] & [0.18 - 0.39] & 17.56 & [0.21 - 1.01] & [0.02 - 0.14] & 44.44\\
\hline
C-Vanderpol & 4 & [0.87 - 1.72] & [0.34 - 0.60] & 17.44 & [0.20 - 0.40] & [0.07 - 0.18] & 44.86\\
% \hline
% LaubLoomis & 7 & [0.003 - 0.016] & [0.18 - 0.49] & 16.78 & [0.001 - 0.003] & [0.11 - 0.30] & 43.50\\
\hline
HybridOsc. & 2 & [0.28 - 0.92] & [0.13 - 0.29] & 16.70 & [0.03 - 0.31] & [0.01 - 0.10] & 45.82\\
\hline
SmoothOsc. & 2 & [0.37 - 1.09] & [0.13- 0.23] & 52.22 & [0.04 - 0.42] & [0.02 - 0.18] & 136.72\\
\hline
Mountain Car & 2 & [0.004 - 0.24] &  [0.08 - 0.22] & 138.90 & [0.0002 - 0.005] & [0.03 - 0.12] & 266.76\\
\hline
Quadrotor & 6 & [0.014 -1.09] & [0.10 - 0.67] & 284.96 & [0.004 - 0.04] & [0.02 - 0.13] & 668.78\\
% \hline
% VehiclePlatoon & 9 & [0.017 - 0.038] & [0.18 - 0.34] & 14.60 & [0.008 - 0.024] & [0.09 - 0.21] & 41.14\\
\hline
\end{tabular}
\caption{\textbf{Evaluations.} The results of \textsf{reachTarget} after  iteration count 1 and 5. We compute average absolute distance $d_a$ and relative distance $d_r$ over 500 iterations of our algorithm for each benchmark. Additionally, a range of values is obtained for $d_a$ and $d_r$ by performing the evaluation on 10 different targets. 
% Both of the Hybrid Oscillator benchmarks originally have 2 modes. The numeral suffix (1 or 2) in their names represents that the corresponding hybrid system has either first mode or second.
}
\label{tab:results}
\end{table}

\begin{figure}
\centering     %%% not \center
\subfigure[Steam]{\label{fig:steam_basic1}\includegraphics[width=70mm]{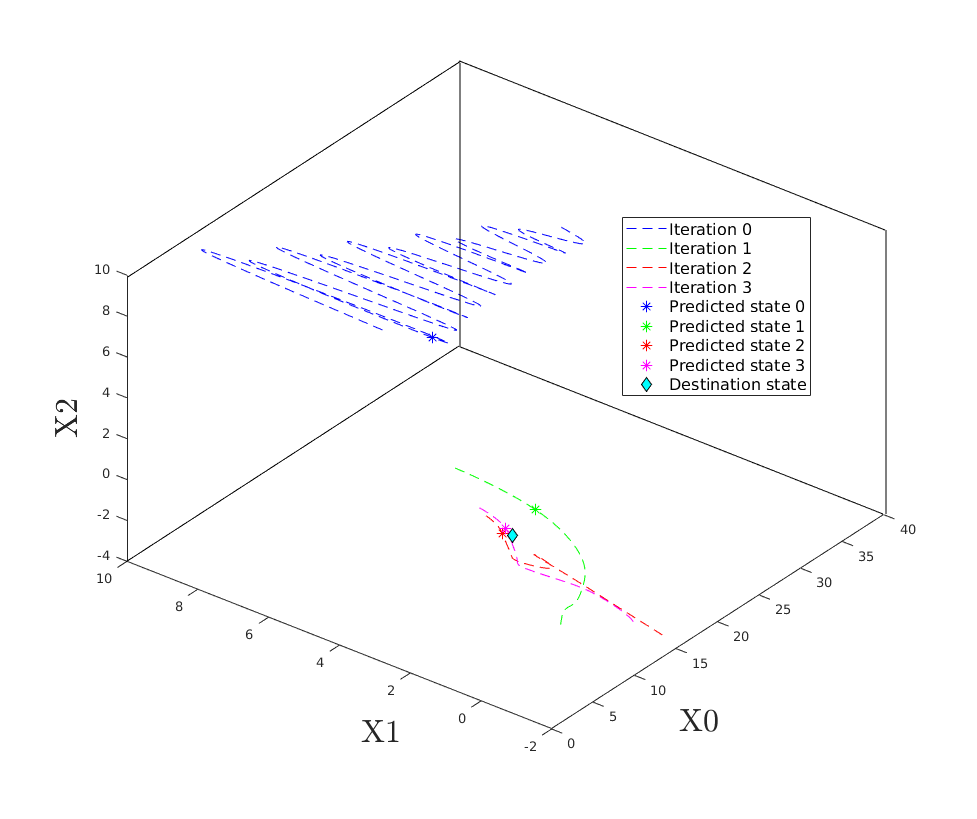}}
\subfigure[Coupled Vanderpol]{\label{fig:coupled_basic1_1}\includegraphics[width=70mm]{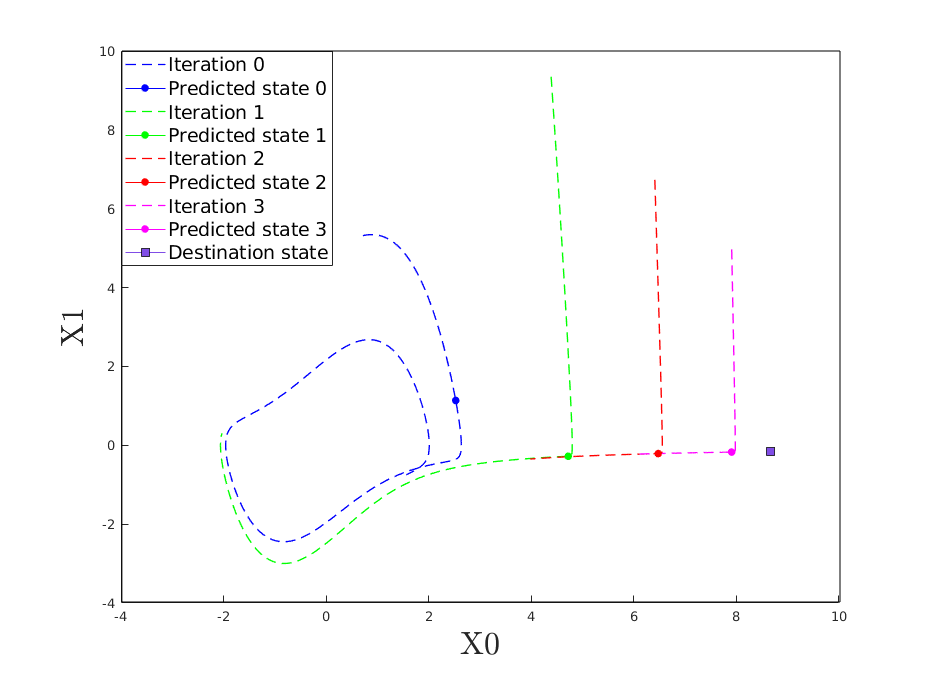}}
\subfigure[Lotka]{\label{fig:lotka_basic1}\includegraphics[width=70mm]{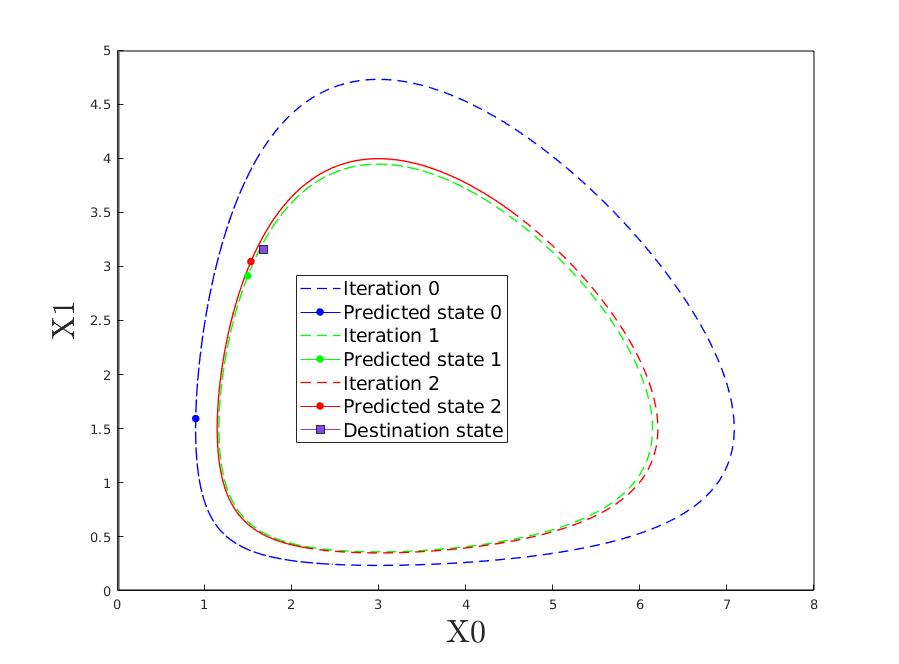}}
\caption{\textbf{Illustration of} \textsf{reachTarget}\textbf{ after iteration count 4}. Iteration 0 is the simulation from $x_{random}$. Subsequent 3 simulations are labeled as Iteration 1, 2 and 3 respectively. As shown, with each iteration, the proximity to the target increases.}
\label{fig:base_approach_exp}
\end{figure}

\begin{figure}
\centering     %%% not \center
\subfigure[Roesseler]{\label{fig:roesseler_basic2}\includegraphics[width=72mm]{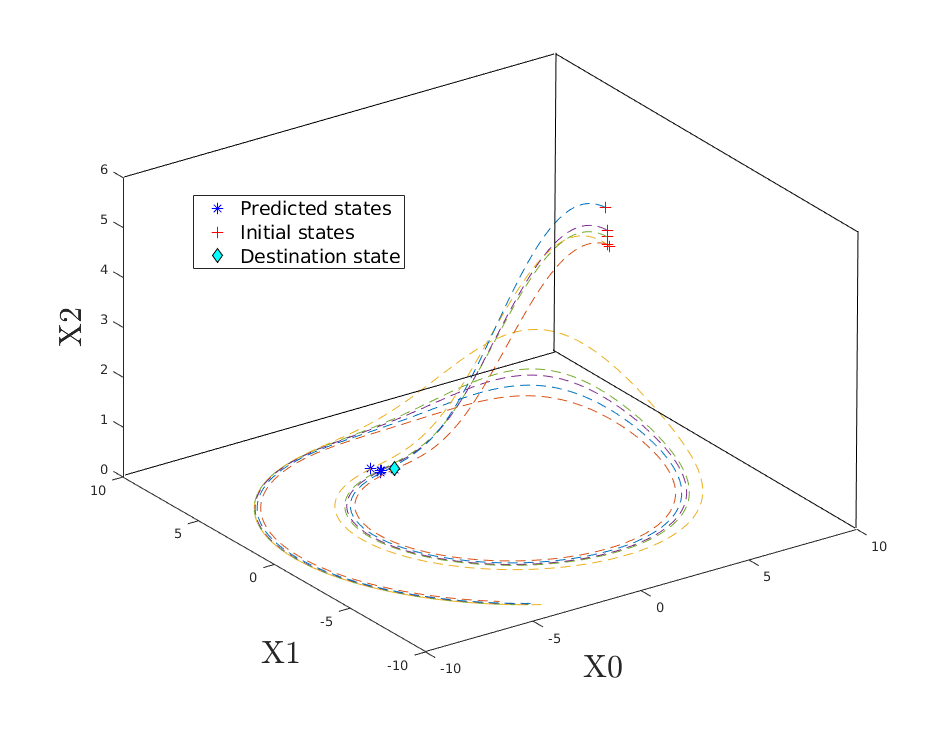}}
\subfigure[Quadrotor]{\label{fig:quad_basic2}\includegraphics[width=72mm]{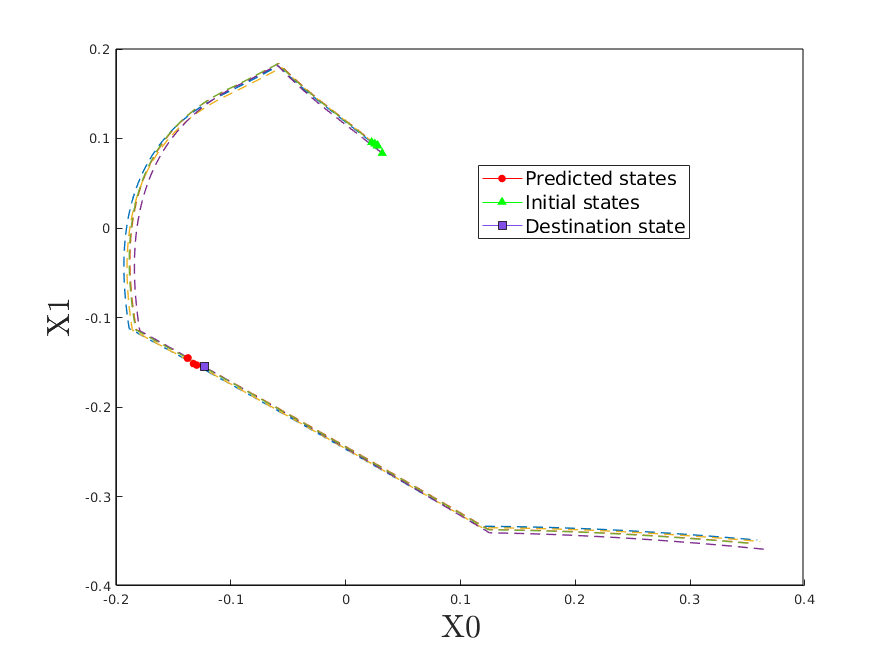}}
\subfigure[Lorentz]{\label{fig:lorentz_basic2}\includegraphics[width=72mm]{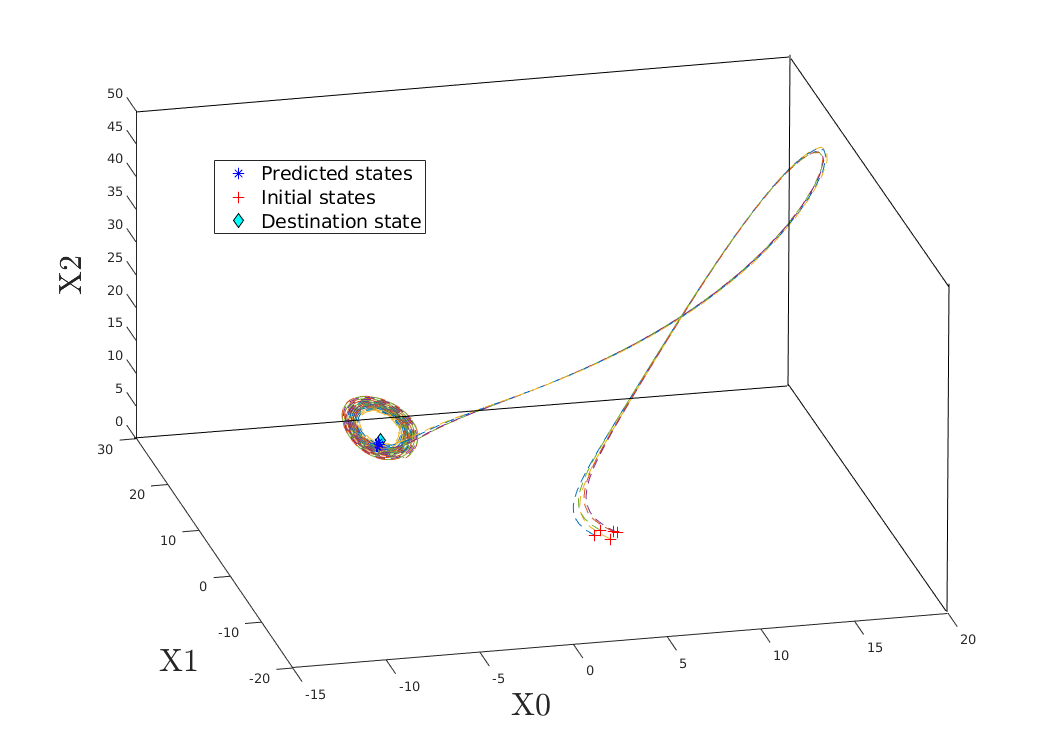}}
\caption{\textbf{Illustration of fixed time instance based evaluation for single target}. Obtaining multiple simulations in the proximity of a given target at a particular time instance.}
\label{fig:fixed_time_exp_single_target}
\end{figure}

\begin{figure}
\centering     %%% not \center
\subfigure[Vanderpol]{\label{fig:vand_mult_dests}\includegraphics[width=72mm]{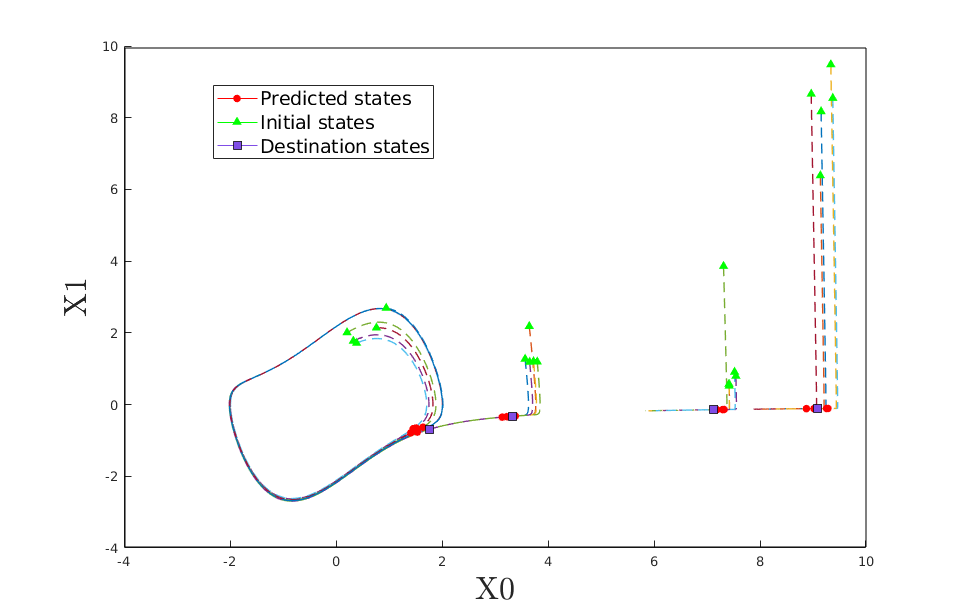}}
\subfigure[Brusselator]{\label{fig:brus_mult_dests}\includegraphics[width=75mm]{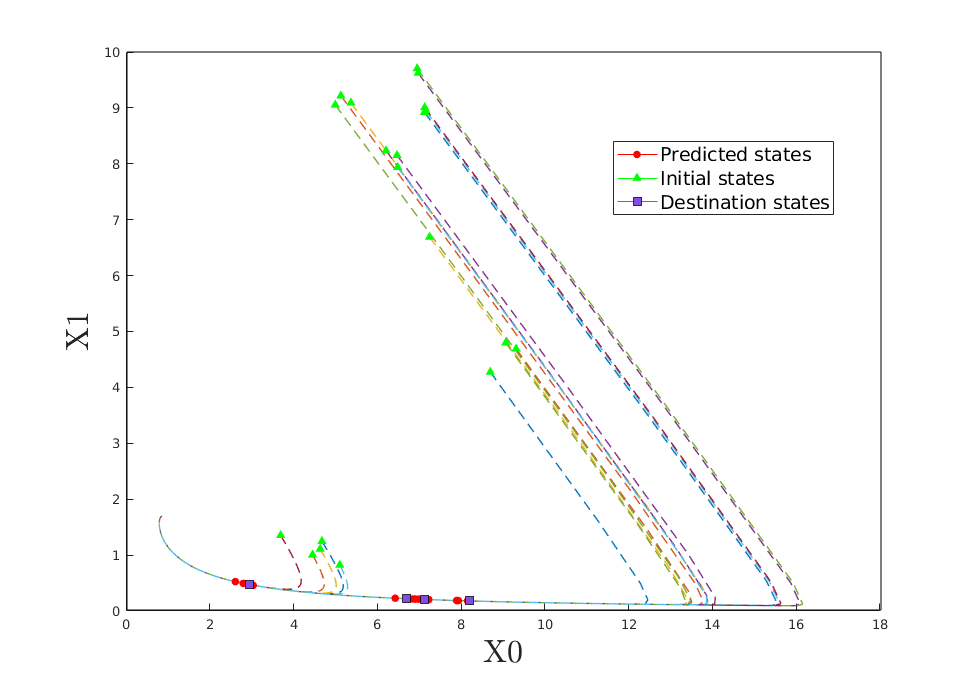}}
\subfigure[Hybrid Linear Oscillator]{\label{fig:hyb_multi_dests}\includegraphics[width=70mm]{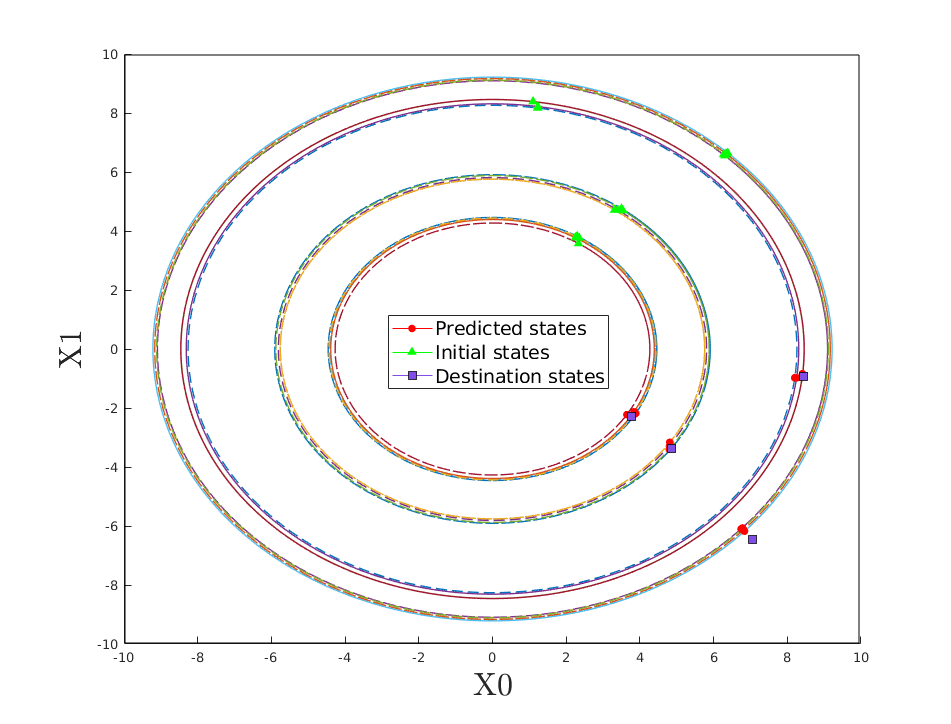}}
\caption{\textbf{Illustration of fixed time instance based evaluation for multiple targets}. Obtaining multiple simulations in the proximity of each target at a particular time instance.}
\label{fig:fixed_time_exp_mult_target}
\end{figure}

We evaluate the performance of \textsf{reachTarget} algorithm as described in Section~\ref{sec:spaceexplore}. We evaluate this by picking a random target state in the domain of interest and let \textsf{reachTarget} generate a trajectory that goes through the neighborhood of the target ($\epsilon = 0.01$ or $0.001$) at a specified time $t$. Typically, \textsf{reachTarget} executes the loop in lines~\ref{ln:begLoop1}-\ref{ln:endLoop1} for 10 times before reaching the target. In Table~\ref{tab:results}, we present the relative and absolute distance between the target and the state reached by the trajectory generated by \textsf{reachTarget} after one or five iterations of the main loop. The demonstration on benchmarks \texttt{Steam}, \texttt{Lotka} and \texttt{Coupled Vanderpol} is shown in Figure~\ref{fig:base_approach_exp}.
Experimental results for the variations of the \textsf{reachTarget} are shown in in Figures~\ref{fig:fixed_time_exp_single_target} and ~\ref{fig:fixed_time_exp_mult_target}

%\footnote{Most of the figures depicting the state space exploration are moved to the appendix due to lack of space.}.

%As explained in Section~\ref{sec:spaceexplore}, our goal is to train the neural network for inverse sensitivity function, generate a random target in the domain, and find a trajectory - using the perturbation obtained from the neural network - that goes through or in the neighborhood around the target. Since the neural network is a function approximator, the trajectory after initial  perturbation need not reach within the bounded threshold of destination. Therefore, we repeat the procedure until a threhold on the number of iterations is reached, or the threshold is satisfied. The evaluation results of running Algorithm~\ref{alg:basicapproach}, \textsf{reachTarget}, for iteration count 1 and 5, respectively, are explained in Table~\ref{tab:results}. 
%The metrics used for evaluation are absolute distance $d_a$ and relative distance $d_r$ defined in Algorithm~\ref{alg:basicapproach}.

\begin{figure}
\centering     %%% not \center
\subfigure[Roesseler: Absolute error]{\label{fig:roesseler_random_abs}\includegraphics[width=65mm]{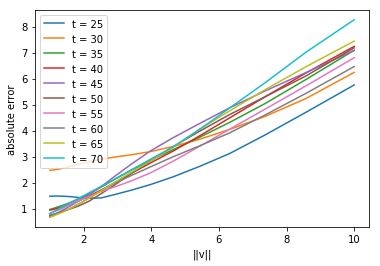}}
\subfigure[Roesseler: Relative error]{\label{fig:roesseler_random_rel}\includegraphics[width=65mm]{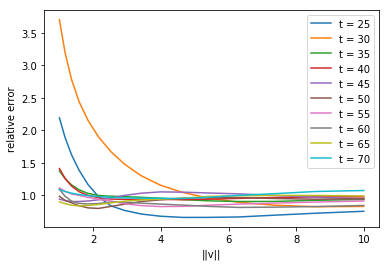}}
\caption{\textbf{Illustration of Random vectors based evaluation.} Plotting absolute and relative errors for random vectors. The horizontal axis denotes the vector norm. Each colored-plot corresponds to a particular time instance. The absolute error increases linearly with vector size, whereas the relative error appears to be converging. Interestingly, we observe similar behavior for each system.}
\label{fig:random_vect_exp}
\end{figure}

We now discuss a few variations of our algorithm and approaches for evaluating it.
%of interesting features of our technique for evaluation.
\begin{enumerate}
\item  \textit{Uncertainty in time:} The control designer might not be interested in reaching the target at a precise time instance as long as it lies within a bounded interval of time. In such cases, one can iterate \textsf{reachTarget} for every step in this interval and generate a trajectory that approaches the target. 
    %For many practical purposes, it may not be possible to know the exact time instance at which the target is reachable from given initial configuration. In such cases, one can provide a time interval, and invoke the algorithm to iteratively find a time instance where the simulation gets reasonably close to the target.

Consider the designer is interested in finding the maximum distance (or, height) the car can go to on the left hill in \texttt{Mountain Car}. By providing an ordered  list of target states and a time interval, she can obtain the maximum distance as well the time instance at which it achieves the maxima (Figure~\ref{fig:mc_time2}). If there is no state in the given initial set from which the car can go to a particular target, the approach, as a side effect, can also provide a suitable initial candidate that takes the car as close as possible to that target. In \texttt{Quadrotor}, one can easily find an initial configuration from which the system can go to a particular location during a given time interval (Fig.~\ref{fig:qr_time}).

\begin{figure}
\centering     %%% not \center
\subfigure[Mountain Car]{\label{fig:mc_time2}\includegraphics[width=90mm]{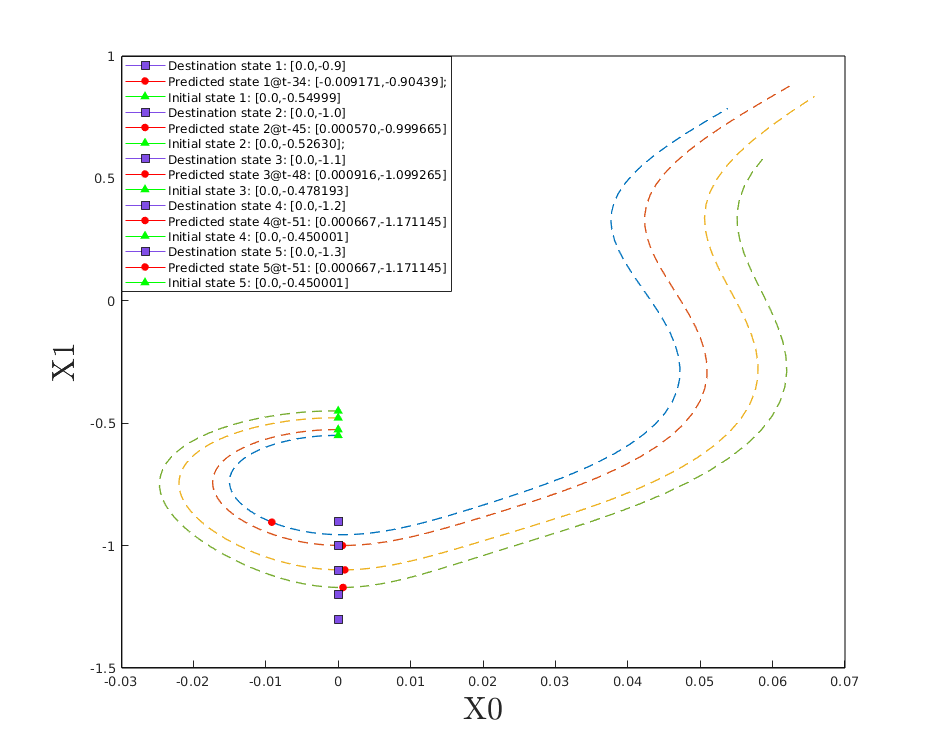}}
\subfigure[Quadrotor]{\label{fig:qr_time}\includegraphics[width=90mm]{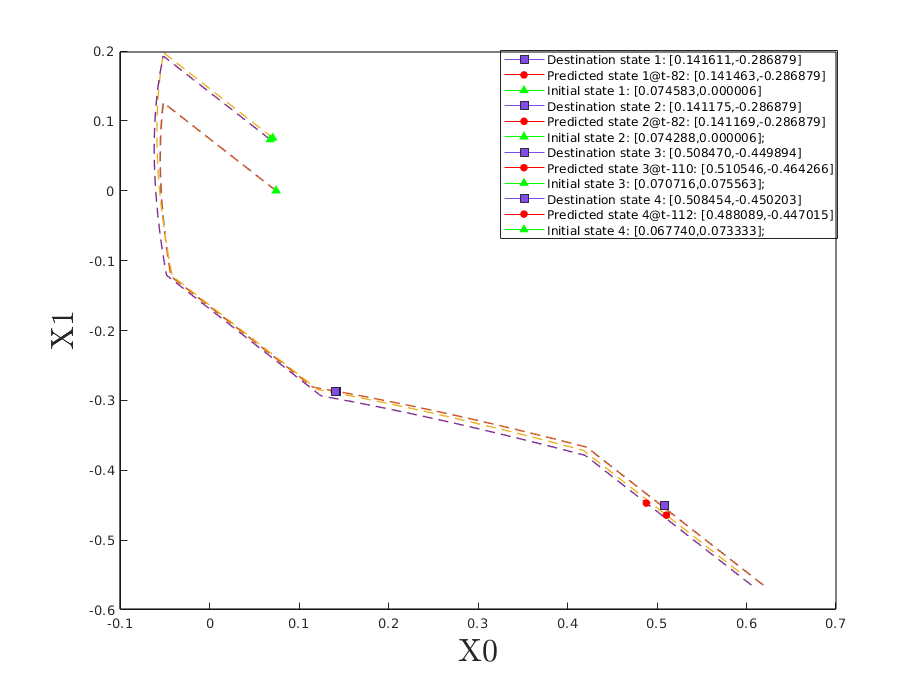}}
\caption{\footnotesize{\textbf{Illustration of time interval based evaluation}. Finding an instance in a given interval at which the distance to the target is minimum. The initial set for \textit{Mountain Car} is [-0.55, -0.45][0.0, 0.0]. \textit{Quadrotor} has 5 discrete modes and each linear segment in~\ref{fig:qr_time} corresponds to a mode.}}
\label{fig:time_interval_exp}
\end{figure}

\item \textit{Generalization:} Based on our \texttt{Mountain Car} experiment, we observed that, for the given initial set, the maximum distance the car can achieve on the left hill is approx. 1.17. However, even after expanding the initial set to [-0.60, -0.40][0.0, 0.0], our approach finds the maximum achievable distance (1.3019) such that the car can still reach on the top of the right hill (shown in Fig.~\ref{fig:mc_largeinitset}). This shows that our neural network is able to generalize the inverse sensitivity over trajectories that go beyond the test cases considered during the training process.

\begin{figure}[!t]
\centering
\includegraphics[width=4.5in]{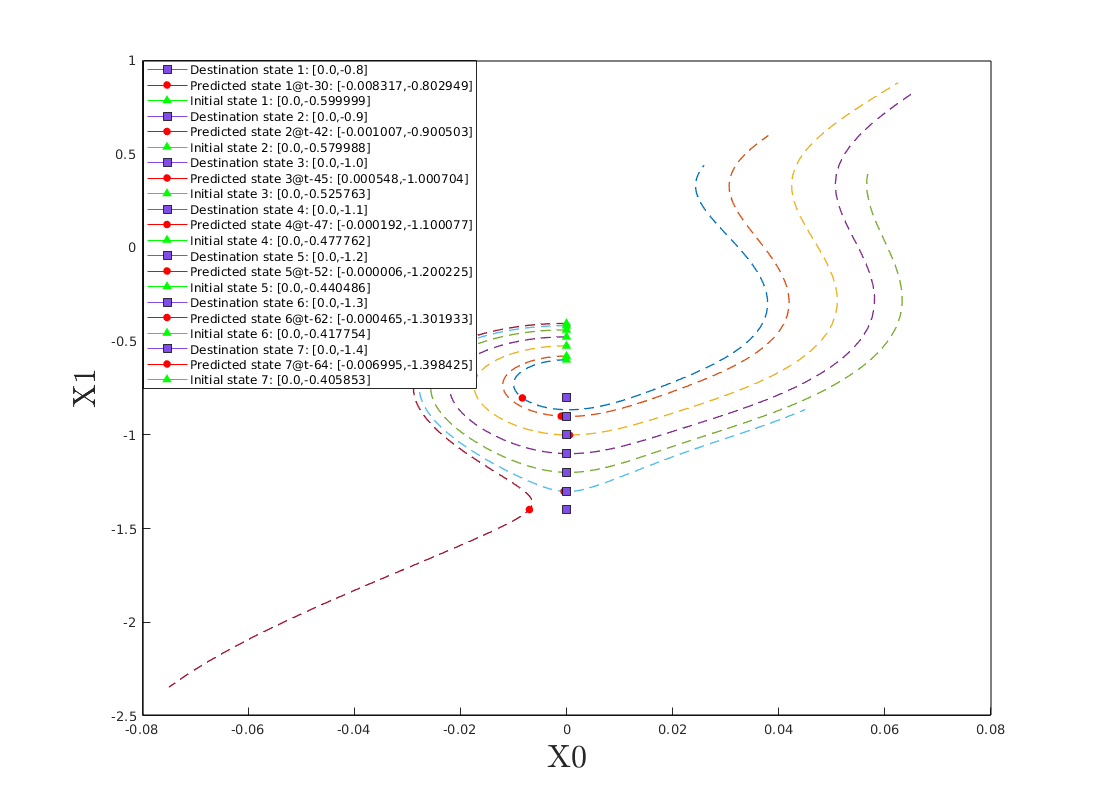}
\caption{\textbf{Illustration of Initial set expansion based evaluation.} Computing the maximum distance the car can achieve on the left hill after expanding the initial set to [-0.60, -0.40][0.0, 0.0].}
\label{fig:mc_largeinitset}
\end{figure}

    %The set of target states can be refined further in order to achieve better precision.
    
\item \textit{Evaluating MRE for Random Targets:} So far we have evaluated our technique with respect to the target states that are reachable from given initial set. We also perform evaluation of \textsf{reachTarget} by generating a random trajectory from the domain and change its course at a provided time interval [25, 70] by a randomly generated vector. The results are demonstrated in Figure~\ref{fig:random_vect_exp}.
% % For the sake of completeness, we also perform evaluation on random proportionally-scaled vectors. We generate a large number of vectors in the unit sphere,  scale them on the scale 1-10, and compute the absolute as well as relative error after first  iteration at time instances in the range . 
\end{enumerate}

%\vspace{-2ex}

\noindent \textbf{Discussion:} It can be observed from Table~\ref{tab:results} that our technique is capable of achieving below 10\% relative distance in almost all cases after 5 iterations. That is, the trajectory generated by \textsf{reachTarget} algorithm after 5 iterations is around 10\% away from the target than the initial trajectory. This was the case even for chaotic systems, hybrid sysems, and for control systems with neural network components. While training the neural network might be time taking process, the average time for generating new trajectories that approach the target is very fast (less than a second for all cases). The high relative distance in some cases might be due to high dimensionality or large distance to the target which may be reduced further with more iterations. 

%The approach can be trivially used to obtain multiple simulations that go within $\epsilon$-neighborhood of a given target. Similarly, one can also provide a set of targets, and for each target, she can obtain multiple trajectories in its proximity. Some of the illustrations are shown 

\subsection{Falsification of Safety Specification}
\label{sec:falsification}

One of the widely used methods for performing state space exploration are falsification methods~\cite{SankaranarayananF12,MCFalsification}. Here, the specification is provided in some temporal logic such as Signal or Metric Temporal Logic~\cite{STLmain,MTLmain}. The falsifier then generates a set of test executions and computes the \emph{robustness} of trajectory with respect to the specification. It then invokes heuristics such as stochastic global optimization for discovering a trajectory that violates the specification.

\begin{figure}
\centering     %%% not \center
\subfigure[Brusselator]{\label{fig:staliro-bruss}\includegraphics[width=60mm]{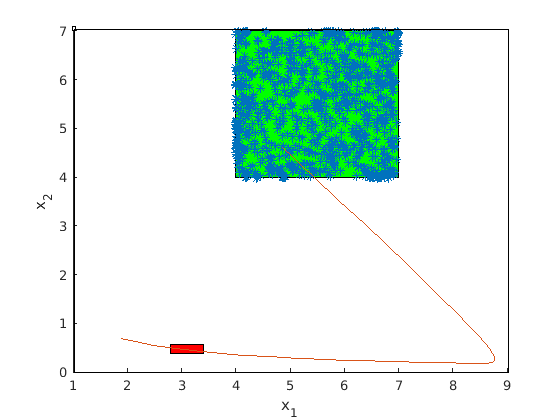}}
\subfigure[Brusselator]{\label{fig:neuralex-bruss}\includegraphics[width=60mm]{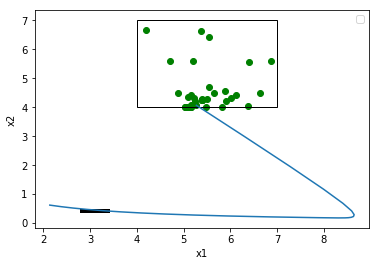}}
\subfigure[Lacoperon]{\label{fig:staliro-lacoperon}\includegraphics[width=60mm]{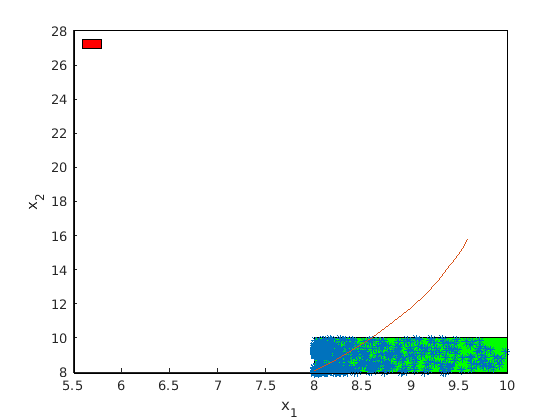}}
\subfigure[Lacoperon]{\label{fig:neuralex-lacoperon}\includegraphics[width=60mm]{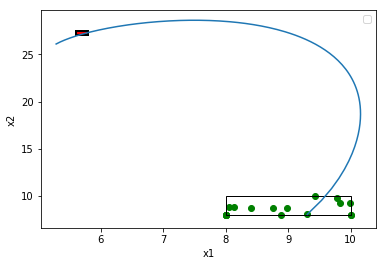}}
\subfigure[Simulated Annealing]{\label{fig:staliro-sa}\includegraphics[width=60mm]{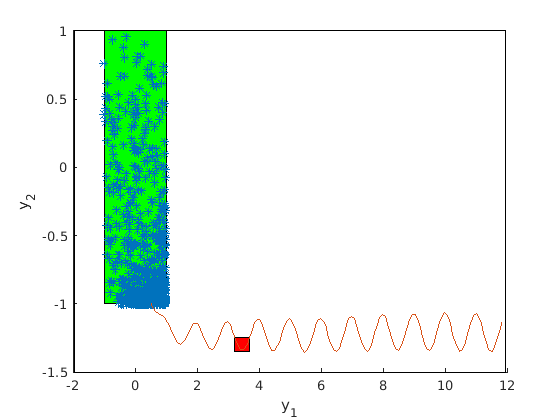}}
\subfigure[Simulated Annealing]{\label{fig:neuralex-sa}\includegraphics[width=60mm]{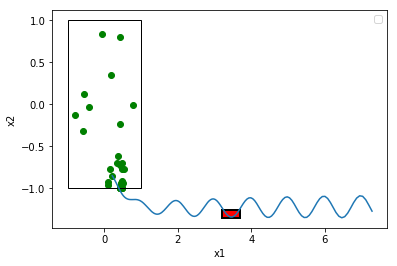}}
\caption{\footnotesize{\textbf{Illustration of falsification in S-Taliro and NeuralExplorer.} When a falsification tool fails to find a counterexample, NeuralExplorer may help (using much less number of samples) the system designer by providing geometric insight into the reason why the property is not satisfied.}}
\label{fig:compare-staliro-bad}
\end{figure}

Given an unsafe set $U$, we provide a simple algorithm to falsify safety specifications. We generate a fixed number ($m$) of random states in the unsafe set $U$. Then, using the \textsf{reachTarget} sub-routine, generate trajectories that reach a vicinity of the randomly generated states in $U$. We terminate the procedure when we discover an execution that enters the unsafe set $U$. For some standard benchmakrs, we compare the number of trajectories generated by STL with the trajectories generated using inverse sensitivity in Fig.~\ref{fig:compare-staliro-bad}. 
The box in each of the figures denotes the initial set and the red box represents the unsafe set. 
Each of the points in the initial set represents a sample trajectory generated by the falsification engine.

\begin{figure}
\centering     %%% not \center
\subfigure[Lorentz]{\label{fig:staliro-lorentz}\includegraphics[width=60mm]{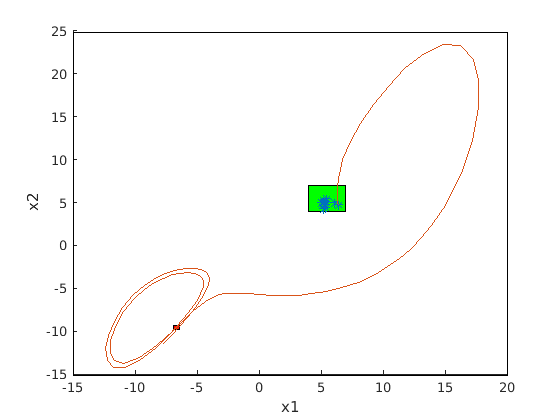}}
\subfigure[Lorentz]{\label{fig:neuralex-lorentz}\includegraphics[width=60mm]{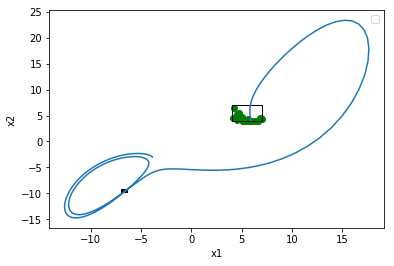}}
\subfigure[Buckling]{\label{fig:staliro-buckling}\includegraphics[width=60mm]{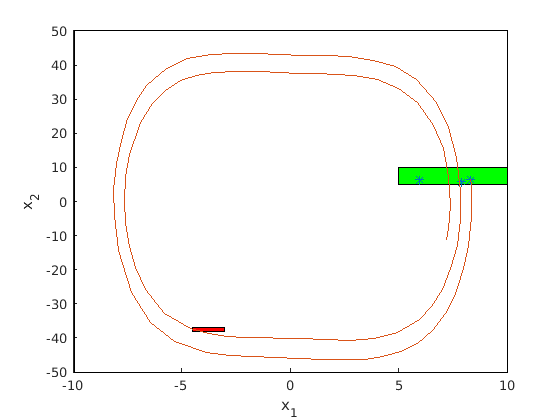}}
\subfigure[Buckling]{\label{fig:neuralex-buckling}\includegraphics[width=60mm]{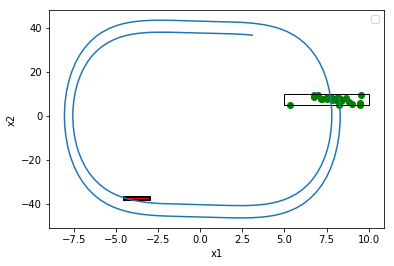}}
\subfigure[Vanderpol]{\label{fig:staliro-vanderpol}\includegraphics[width=60mm]{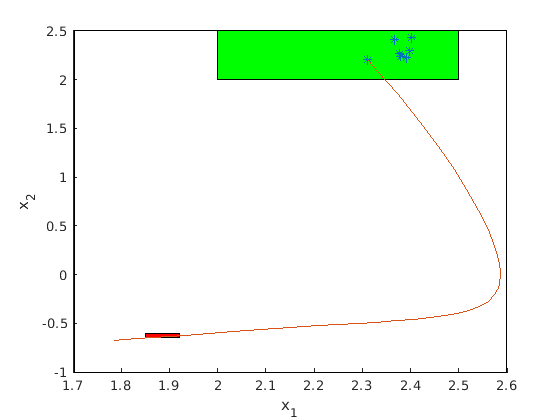}}
\subfigure[Vanderpol]{\label{fig:neuralex-vanderpol}\includegraphics[width=60mm]{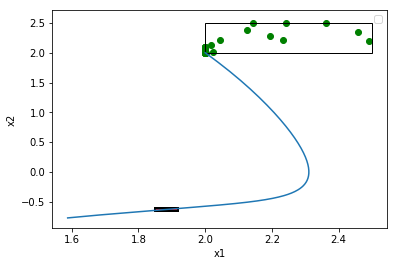}}
\caption{\textbf{Illustration of falsification in S-Taliro and NeuralExplorer.} Some scenarios where S-Taliro terminates with a falsification trajectory faster than our approach.}
\label{fig:compare-staliro-good}
\end{figure}

Falsification using approximation of inverse sensitivity enjoys a few advantages over other falsification methods. 
First, since our approach approximates the inverse sensitivity, and we use the \textsf{reachTarget} sub-routine; if the approximation is accurate to a certain degree, each subsequent trajectory generated in \textsf{reachTarget} would make progress towards the destination. 
Second, if the safety specification is changed slightly the robustness of the trajectories with respect to new specification and the internal representation for the stochastic optimization solver has to be completely recomputed. 
However, since our trajectory generation does not rely on computing the robustness for all the previously generated samples, our algorithm is effective even when the safety specification is modified.

The third and crucial advantage of our approach lies when the falsification tool does not yield a counterexample. 
In those cases, the typical falsification tools cannot provide any geometric insight into the reason why the property is not satisfied.
However, using an approximation of inverse sensitivity, the system designer can envison the required perturbation of the reachable set in order to move the trajectory in a specific direction.
This geometric insight would be helpful in understanding why a specific trajectory does not go into the unsafe set.

Considering these advantages, the results demonstrated in Fig.~\ref{fig:compare-staliro-bad} should not be surprising.
We also would like to mention that these advantages come at the computational price of training the neural networks to approximating the inverse sensitivity. 
In addition to the examples shown above, we have included some examples (see Fig.~\ref{fig:compare-staliro-good}) where S-Taliro terminates with a falsification trajectory faster than our approach. The reasons for such cases and methods to improve falsification using NeuralExplore are a topic of future work.

\subsection{Density Based Search Methods For State Space Exploration}
\label{sec:density}

One of the most commonly used technique for performing state space exploration is generation of trajectories from a set of random points generated using an apriori distribution.
Based on the proximity of these trajectories to the \emph{unsafe} set, this probability distribution can further be refined to obtain trajectories that move closer to the unsafe set.
However, one of the computational bottlenecks for this is the generation of trajectories.
Since the numerical ODE solvers are sequential in nature, the refinement procedure for probability distribution is hard to accelerate.

\begin{figure}
\centering     %%% not \center
\subfigure[]{\label{fig:reach-fwd-vander-pc}\includegraphics[width=70mm]{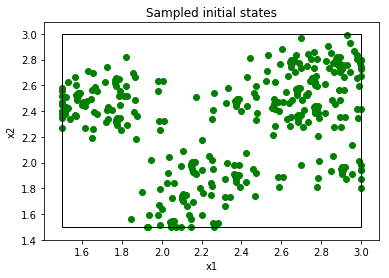}}
\subfigure[]{\label{fig:reach-fwd-vander-proj-trajs}\includegraphics[width=60mm]{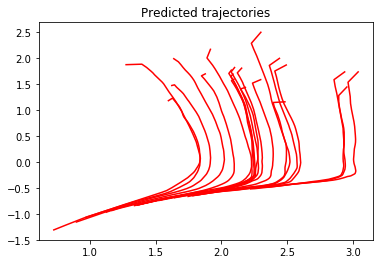}}
\subfigure[]{\label{fig:reach-fwd-vander-act-trajs}\includegraphics[width=60mm]{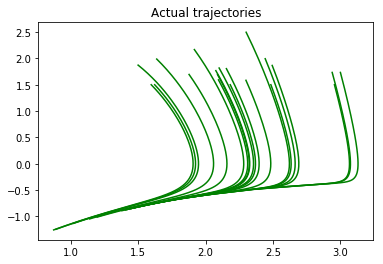}}
\subfigure[]{\label{fig:reach-fwd-vander-proj-disp}\includegraphics[width=60mm]{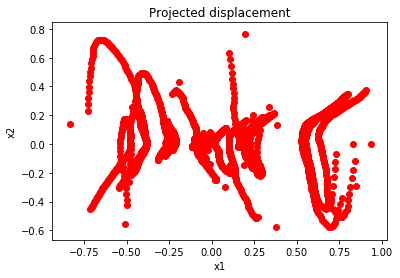}}
\subfigure[]{\label{fig:reach-fwd-vander-act-disp}\includegraphics[width=60mm]{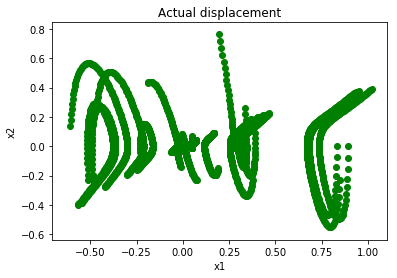}}
\caption{\footnotesize{\textbf{Illustration of sensitivity in \textit{reachability} using \texttt{Vanderpol}}. A cluster of points is sampled in the neighborhood of a reference state. Actual trajectories as well as predicted trajectories obtained by the neural network which approximates sensitivity function are shown. Their corresponding displacement from the reference trajectory at each time step are also displayed in (d) and (e).}}
\label{fig:reach-forward-vanderpol}
\end{figure}

\begin{figure}
\centering     %%% not \center
\subfigure[]{\label{fig:reach-fwd-jetengine-pc}\includegraphics[width=70mm]{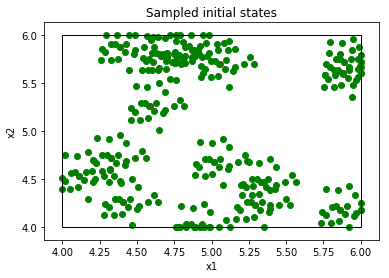}}
\subfigure[]{\label{fig:reach-fwd-jetengine-proj-trajs}\includegraphics[width=60mm]{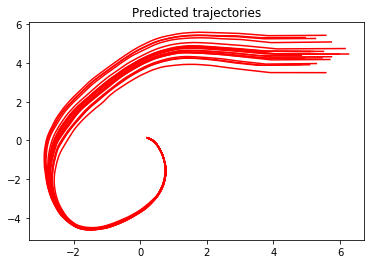}}
\subfigure[]{\label{fig:reach-fwd-jetengine-act-trajs}\includegraphics[width=60mm]{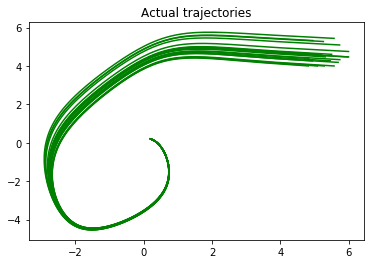}}
\subfigure[]{\label{fig:reach-fwd-jetengine-proj-disp}\includegraphics[width=60mm]{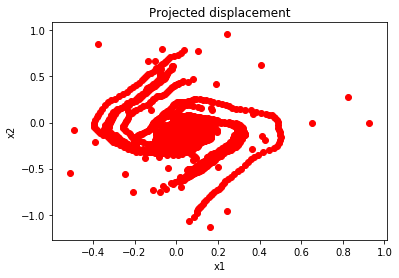}}
\subfigure[]{\label{fig:reach-fwd-jetengine-act-disp}\includegraphics[width=60mm]{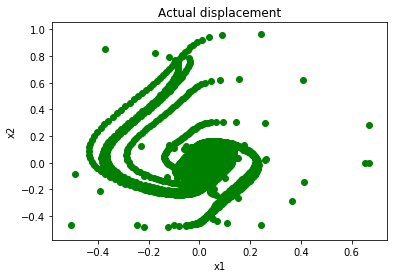}}
\caption{\textbf{Illustration of sensitivity in \textit{reachability}  using \texttt{Jetengine}}. Actual trajectories as well as projected trajectories for a cluster of points sampled in the neighborhood of a reference trajectory are shown. In addition, their displacement from the reference trajectory at each time instance are displayed.}
\label{fig:reach-forward-jetengine}
\end{figure}

For this purpose, one can use the neural network approximation of sensitivity to \emph{predict} many trajectories in an embarassingly parallel way.
Here, a specific set of initial states for the trajectories are generated using a pre-determined distribution.
Only a few of the corresponding trajectories for the initial states are generated using numerical ODE solvers. 
These are called as \emph{anchor trajectories}.
The remainder of trajectories are not generated, but rather predicted using the neural network approximation of sensitivity and anchor trajectories. That is, $\xi(x_i,t) + \Phi_{NN}(x_i, x_j - x_i, t)$.
Additionally, the designer has the freedom to choose only a subset of the initial states for only a specific time interval for prediction and refine the probability distribution for generating new states.
This would also allow us to specifically focus on a time interval or a trajectory without generating the prefix of it.
A few examples of predictive trajectory generation for performing \emph{reachability analysis} on \texttt{Vanderpol oscillator} and \texttt{Jetengine} are provided in Fig.~\ref{fig:reach-forward-vanderpol} and Fig.~\ref{fig:reach-forward-jetengine}.

% \vspace{-5ex}

\subsubsection{Density Based Search for Falsification}

Similar to the inverse sensitivity based falsification, one can use the density based search space method for generating trajectories that reach a destination and violate a safety specification.
The forward density based search procedure would work as follows.
First, an anchor trajectory is generated and time intervals of its trajectory that are closer to the unsafe set are identified.
Then a set of new initial states are generated according to an apriori decided distribution.
Instead of generating the trajectory from these initial states, the \emph{predicted trajectories} using the anchor trajectory and neural network approximation of sensitivity is generated specifically for the time intervals of interest.
Then, the initial state with the predicted trajectory that is closest to the unsafe set is chosen and a new anchor trajectory from the selected intial state is generated.
This process of generating anchor trajectory, new distribution of initial states, and moving closer to the unsafe set is continued until you reach within the threhold that is generated by the user.

\begin{figure}
\centering     %%% not \center
\subfigure[]{\label{fig:false-fwd-vander-proj-pc}\includegraphics[width=60mm]{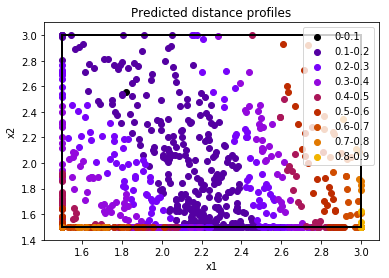}}
\subfigure[]{\label{fig:false-fwd-vander-act-pc}\includegraphics[width=60mm]{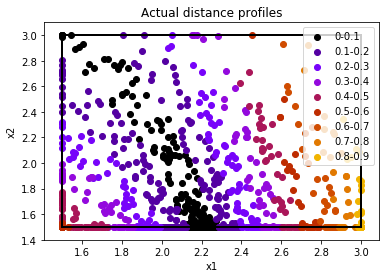}}
\subfigure[]{\label{fig:false-fwd-vander-proj-disp1}\includegraphics[width=60mm]{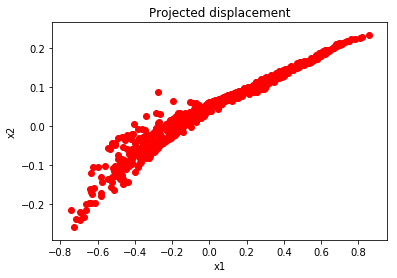}}
\subfigure[]{\label{fig:false-fwd-vander-act-disp1}\includegraphics[width=60mm]{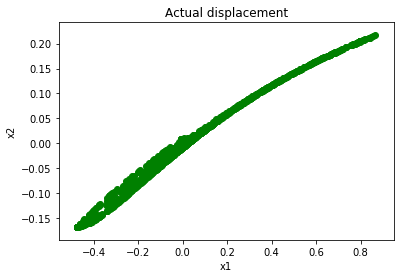}}
\caption{\footnotesize{\textbf{Illustration of sensitivity in \textit{falsification} using \texttt{Vanderpol}.} The perturbation in the neighborhood of reference state are greedily chosen in an iterative manner so as to minimize the distance to unsafe state. The projected and actual displacements from reference trajectories are shown. The sampled states are classified based on their euclidean distance to the unsafe state.}}
\label{fig:falsify-forward-vanderpol}
\end{figure}

\begin{figure}
\centering     %%% not \center
\subfigure[]{\label{fig:false-fwd-bruss-proj-pc}\includegraphics[width=60mm]{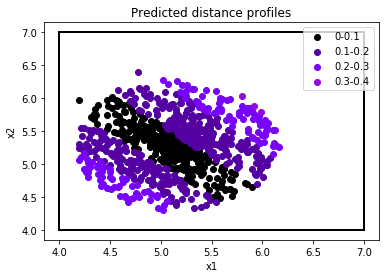}}
\subfigure[]{\label{fig:false-fwd-bruss-act-pc}\includegraphics[width=60mm]{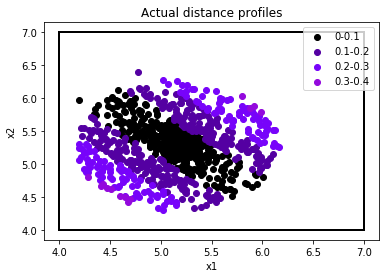}}
\subfigure[]{\label{fig:false-fwd-bruss-proj-disp1}\includegraphics[width=60mm]{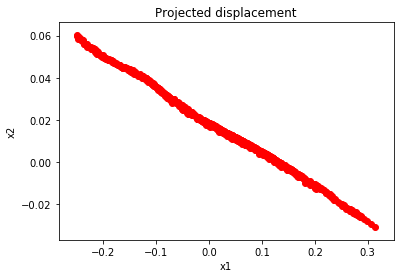}}
\subfigure[]{\label{fig:false-fwd-bruss-act-disp1}\includegraphics[width=60mm]{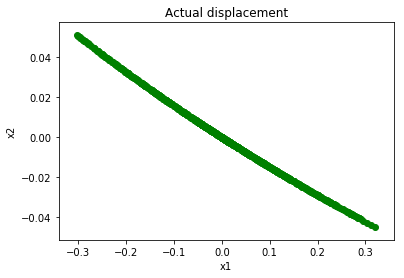}}
\caption{\textbf{Illustration of sensitivity in \textit{falsification}  using \texttt{Brusselator}}. For a given unsafe state, a reference initial state and a set of perturbations in its neighborhood are randomly chosen. The perturbation that minimizes the distance to the unsafe state is greedily picked up and the new reference trajectory is sampled from this perturbation. The procedure repeats until the threshold is reached or the iteration count is exhausted. The states explored in this iterative process are classified (in Fig.~\ref{fig:false-fwd-bruss-proj-pc} and ~\ref{fig:false-fwd-bruss-act-pc}) based on their respective distance to the unsafe state.}
\label{fig:falsify-forward-bruss}
\end{figure}

\begin{figure}
\centering     %%% not \center
\subfigure[Brusselator: Distance profiles]{\label{fig:false-inv-bruss}\includegraphics[width=60mm]{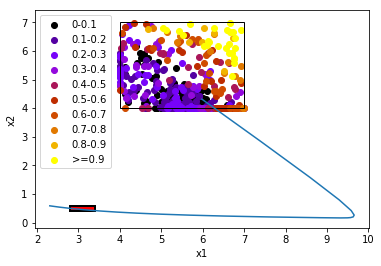}}
\subfigure[Brusslator: Distance profiles for a different unsafe spec]{\label{fig:false-inv-bruss-diff-usafe}\includegraphics[width=60mm]{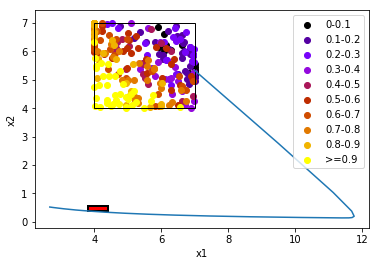}}
\subfigure[Vanderpol: Distance profiles]{\label{fig:false-inv-vander}\includegraphics[width=60mm]{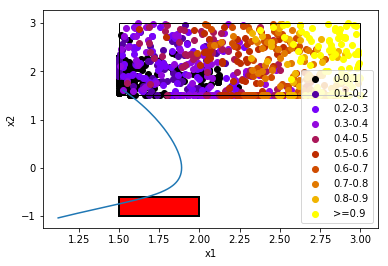}}
\subfigure[Vanderpol: Distance profiles for a different time instance]{\label{fig:false-inv-vander-timechange}\includegraphics[width=60mm]{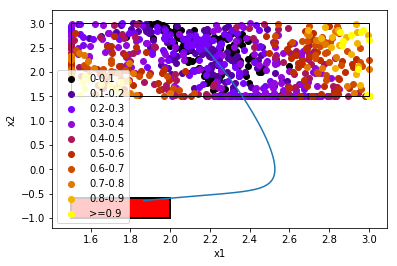}}
\caption{\textbf{Illustration of inverse sensitivity in \textit{falsification}.} Given a set of states satisfying unsafe spec and a time instance, the routine, starting from a reference trajectory,  attempts to iteratively find a falsifying trajectory. The initial states explored in the process are colored according to their distance to unsafe states. These color densities help in identifying regions in the initial set potentially useful for falsification. Notice the difference in the distance profiles (color densities) as we change the time instance (in \texttt{Vanderpol}) or select a difference unsafe spec (in \texttt{Brusselator}).}
\label{fig:falsify-inverse}
\end{figure}

Demonstration of this procedure for \texttt{Vanderpol} and \texttt{Brusselator} systems are shown in Fig.~\ref{fig:falsify-forward-vanderpol} and ~\ref{fig:falsify-forward-bruss} respectively. Notice that this approach gives an underlying intuition about the geometric behavior of neighboring trajectories. A similar method for density based estimation using inverse sensitivity approximation can also be devised. Instead of sampling the initial set, the density based method for inverse sensitivity generates random states around the unsafe set to be reached and then generates a density map. An example of such a density map generated is given in Fig.~\ref{fig:falsify-inverse}.
%Due to the lack of space, we do not fully describe it here. 

\section{Conclusion and Future Work}
We presented NeuralExplorer framework for state space exploration of closed loop control systems using neural network. Our framework depends on computing neural network approximations of two key properties of a dynamical system called sensitivity and inverse sensitivity. We have demonstrated that for standard benchmarks, these functions can be learned with less than $20\%$ relative error. We demonstrated that our method can not only be applied to standard nonlinear dynamical systems but also for control systems with neural network as feedback functions.

Using these approximations of sensitivity and inverse sensitivity, we presented new ways to performing state space exploration. We also highlighted the advantages of the falsification methods devised using the approximations. Additionally, we demonstrated that our techniques give a geometric insight into the behavior of the system and provide more intuitive information to the user, unlike earlier black box methods. We believe that these techniques can help the system designer in search of the desired executions.~\footnote{All the code and examples for the evaluations performed in this section are available at~\url{https://github.com/mag16154/NeuralExplorer
}}
%
% Based on the evaluations, we were able to generate trajectories that reach a target state or its close neighborhood for a wide variety of systems. 

In future, we intend to extend this work to handle more generic systems such as feedback systems with environmental inputs. We believe such a black-box method for generating adversarial examples can be integrated into generative adversarial training for training neural networks for control applications. 
%The high relative distance in some cases might be due to high dimensionality or large distance to the target. For the future work, we are interested in extending the technique to systems with time variant control inputs.
% Acknowledgments---Will not appear in anonymized version
%\acks{We thank a bunch of people.}

\bibliographystyle{splncs04}
\bibliography{main}
\end{document}